\documentclass[12pt,draftclsnofoot, onecolumn]{IEEEtran}
\usepackage{graphicx,cite,epsfig,amssymb,amsmath,url,stfloats,latexsym,threeparttable,balance}
\usepackage{multirow}
\usepackage{color,epstopdf}
\usepackage{booktabs}
\usepackage{setspace}
\usepackage{amsbsy}
\usepackage{cite}
\usepackage{mathrsfs}
\usepackage{amsfonts}
\usepackage{subfig}

\begin{document}
\markboth{IEEE Wireless Communications, Vol. XX, No. YY, Month 2024}
{Chen, Zhao, Chen, Lu, \& Meng: A Discrete Packet Collision Model \ldots}
\title{\mbox{}\vspace{1.5cm}\\
\textsc{RIS-Based Over-the-Air Diffractional Channel Coding} \vspace{1.5cm}}
\author{Yingzhe Hui, Shuyi Chen{$^{^\dagger}$},~\IEEEmembership{Member,~IEEE}, Yifan Qin, Weixiao Meng,~\IEEEmembership{Senior Member,~IEEE}, Qiushi Zhang and Wei Jin %
\thanks{Yingzhe Hui (e-mail: {\tt yingzhe\_hui@stu.hit.edu.cn}), Shuyi Chen (e-mail: {\tt chenshuyitina@gmail.com}), Weixiao Meng (e-mail: {\tt wxmeng@hit.edu.cn})  (the corresponding Author) and Qiushi Zhang (e-mail: {\tt 21b905030@stu.hit.edu.cn}) are with the Communication Research Center, Harbin Institute of Technology, China. Yifan Qin (e-mail: {\tt qinyifan@hrbeu.edu.cn}) and Wei Jin (e-mail: {\tt hrbeu\_jinwei@163.com}) are with the Key Lab of In-fiber Integrated Optics, Ministry Education of China, Harbin Engineering University, China.}
\thanks{This work was supported in part by the National Science Fund for Young Scholars No. 62201176, Natural Science Foundation of Heilongjiang Province of China No. YQ2023005, Young Elite Scientist Sponsorship Program by CAST No. YESS20210339.}
\thanks{The paper was submitted \today.}\\
\vspace{1.5cm}
\underline{{$^{^\dagger}$}Corresponding Author's Address:}\\
$\mbox{Shuyi~Chen}$\\
 Communication Research Center\\
Harbin Institute of Technology, China\\
Tel: +86-18645080001\\
Email: {\tt chenshuyi@hit.edu.cn}}

\date{\today}
\renewcommand{\baselinestretch}{1.2}
\thispagestyle{empty} \maketitle \thispagestyle{empty}
\newpage
\setcounter{page}{1}

\begin{abstract}
Reconfigurable Intelligent Surfaces (RIS) are programmable metasurfaces utilizing sub-wavelength meta-atoms and a controller for precise electromagnetic wave manipulation. This work introduces an innovative channel coding scheme, termed RIS-based diffractional channel coding (DCC), which capitalizes on diffraction between two RIS layers for signal-level encoding. Contrary to traditional methods, DCC expands signal dimensions through diffraction, presenting a novel countermeasure to channel effects. This paper focuses on the operational principles of DCC, including encoder and decoder designs, and explores its possibilities to construct block and trellis codes, demonstrating its potential as both an alternative and a supplementary conventional coding scheme. Key advantages of DCC include eliminating extra power requirements for encoding, achieving computation at the speed of light, and enabling adjustable code distance, making it a progressive solution for efficient wireless communication, particularly in systems with large-scale data or massive MIMO.
\end{abstract}
\begin{IEEEkeywords}
\centering
Reconfigurable intelligent surface (RIS), diffractional channel coding (DCC), signal-level error control coding, multi-layer RIS
\end{IEEEkeywords}

\IEEEpeerreviewmaketitle

\vspace{0.4in}
\section{Introduction}
The flexible manipulation of electromagnetic waves holds paramount significance in modern wireless communication, offering substantial gains and benefits. Meanwhile, the emergence of programmable metasurfaces offers unprecedented opportunities for control and manipulation of electromagnetic properties, serving as a catalyst for designing more efficient communication paradigms. Reconfigurable intelligent surfaces (RIS), an example of programmable metasurfaces, employ sub-wavelength meta-atoms in combination with a controller to manipulate electromagnetic waves with great precision. Researchers can dynamically adjust the electrical parameters of the meta-atoms in real-time to achieve the desired results, leading to the development of advanced applications in wireless communication, sensing, imaging, and other fields~\cite{Meta2022CuiNature,Hua2024}.

The use of RIS has revolutionized approaches to wireless communication. In addition to enhancing wireless signal strength and coverage, RIS can serve as an alternative to specific conventional modules in transmitters and receivers. For example, RIS can perform modulation by inputting the data source into the controller and mapping the data source into signals with different amplitudes and phases~\cite{Li2021MIMOWC}. Furthermore, RIS supports multi-antenna transmission by using each meta-atom as a transmit antenna and inputting the precoding matrix into the controller to achieve analog beamforming~\cite{Lu2022PrecodingICN}.

However, the potential of RIS extends beyond previous assumptions. Recent studies have shown that multi-layer RIS can achieve a deep neuron network and support over-the-air signal processing tasks such as object recognition and image reconstruction~\cite{Chen2024, Meta2022CuiNature}. With the advent of multi-layer RIS, it is possible to perform complex signal processing tasks without relying on traditional hardware, which could lead to cost savings and a reduction in energy consumption.

Inspired by the signal processing capability of multi-layer RIS, in this paper, we aim to exploit the innovative use of multi-layer RIS to enhance error control coding by harnessing the propagation between two layers of RIS. As the propagation between two adjacent RIS layers is modeled using the Rayleigh-Sommerfeld diffraction equation~\cite{ Mengu2022MultilayerSR}, we can obtain the transformation matrix between the signal after the first layer and the signal before the second layer. The Rayleigh-Sommerfeld equation, a rigorous solution, precisely satisfies Maxwell's and Helmholtz's equations. It seamlessly integrates the Huygens principle and accurately describes coherent light propagation in diffracted fields. When the first layer contains fewer meta-atoms than the second, the dimension of the output signal, or the signal received by the second layer, increases. This increment introduces an additional signal, essentially a linear combination of the original signal. This enables a form of channel coding via diffraction between the two RIS layers, which we term RIS-based diffractional channel coding (DCC). DCC is expected to inherit the benefits brought by multi-layer RIS, i.e., light-speed computation, energy conservation, scalability, and compatibility with massive MIMO, which makes it a potential substitute or supplement of the conventional coding scheme. Meanwhile, we also want to point out that the proposed RIS-DCC scheme shares the basic idea with signal-level coding schemes, such as physical-layer network coding (PNC), which exploits the inherent additivity of simultaneously arriving electromagnetic waves for equivalent coding operations~\cite{Chen2020PNCNetwork}. The key difference is that the equivalent coding operation of PNC occurs at the receiver, while RIS-DCC occurs at the transmitter. While existing signal-level coding methods provide a foundation for our system design and signal processing algorithms, applying RIS to a channel coding operation introduces novel challenges, marking a pioneering approach in this field.

In conclusion, this vision paper aims to introduce a signal-level over-the-air RIS-based coding scheme, namely, DCC, offering a potential new paradigm in wireless communications. The rest of this paper is outlined as follows. We first provide essential background knowledge on RIS-based applications. Subsequently, the DCC scheme is detailed by depicting the transceiver model and fundamental concepts such as the generator matrix and code distance. To provide a comprehensive insight, we exemplify the capabilities of (4,2) RIS-DCC in constructing block and trellis codes. Then, to show the performance of DCC, two representative block and trellis codes, i.e., (7,4) Hamming code and (2,1,3) convolutional code, are chosen and compared with equivalent DCC structures. Additionally, we discuss the benefits and limitations of the proposed scheme, followed by the conclusions.

\section{Background knowledge}
\begin{figure}
\centering
\includegraphics[width=14cm]{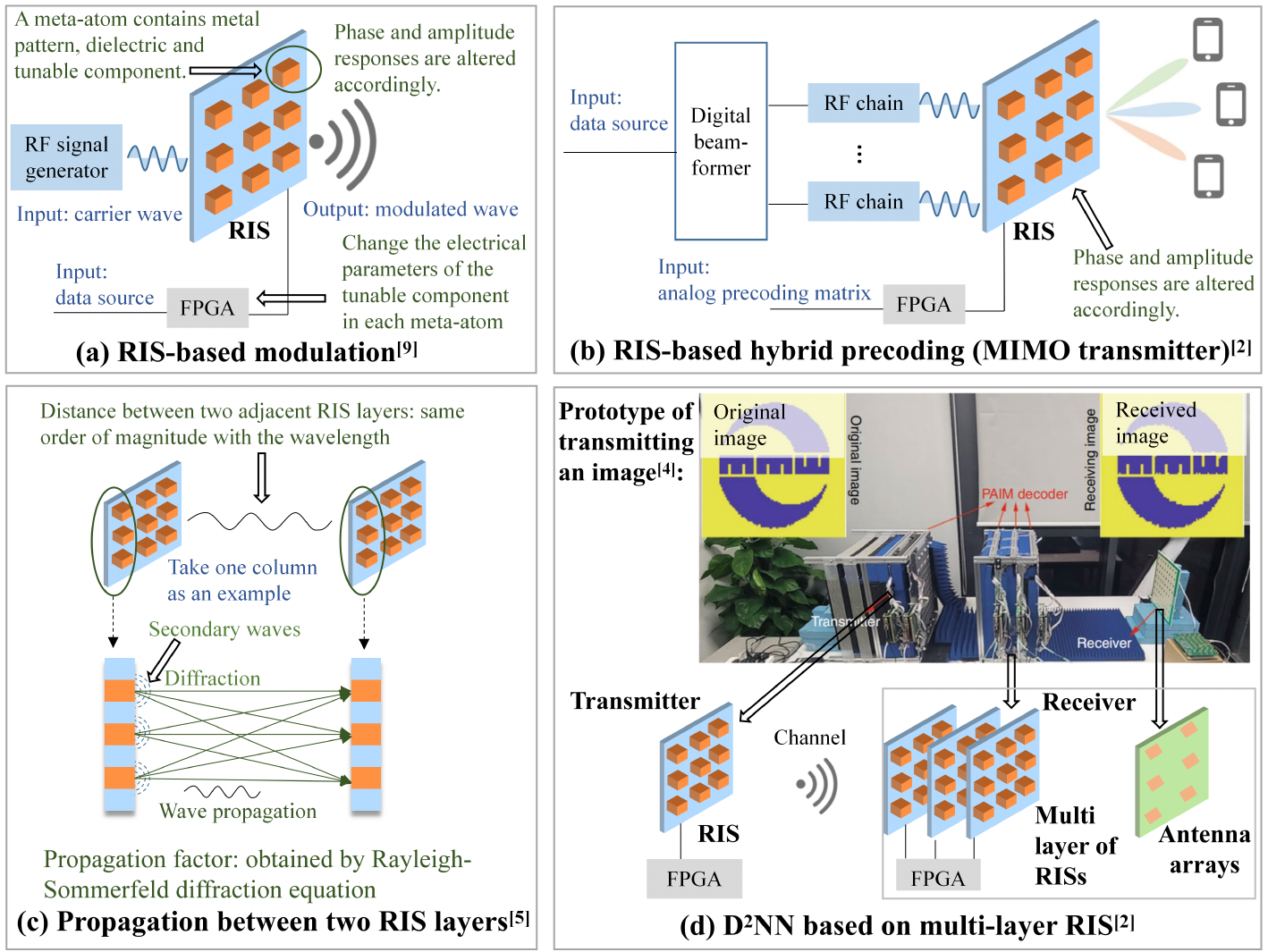}
\caption{Background knowledge, where (a) (b) show the schematic diagrams of RIS-based modulation and MIMO transmitter, (c) explains the diffractional propagation between two RIS layers, and (d) uses a prototype of image transmission to show the potentialities of using multi-layer RIS. }
\label{background}
\end{figure}
\subsection{RIS-based modulation}
RIS uses programmable meta-atoms, controlled via FPGA, to modulate signals by dynamically altering electromagnetic wave properties. This capability facilitates the direct mapping of data into signals with tailored amplitudes and phases, allowing transmission of bitstreams equal to the meta-atom count~\cite{Wu2021ModuTVT,Li2021MIMOWC}. Compared to traditional wireless communication transmitter architectures, RIS-based modulation offers advantages such as lower hardware complexity, simpler architecture, reduced power consumption, and easier manufacturing. Despite the technical challenge of coupling between amplitude and phase adjustments, RIS supports various modulation schemes, with a focus on constant envelope modulations like QPSK and 8PSK. Innovations in non-linear modulation have enabled high-speed, robust QAM modulation with a 20 Mbps data rate~\cite{Tang2020RISJSAC}.

\subsection{RIS-based MIMO transmitter}
By using a precoding matrix to control the response of each meta-atom, RIS enables analog beamforming~\cite{Deng2022RHSTWC}, effectively serving as a MIMO transmitter~\cite{Lu2022PrecodingICN}, as illustrated in Fig.~\ref{background}(b). The integration of RIS in MIMO systems, by minimizing the reliance on traditional RF chains, presents a compelling solution for low-cost, low power consumption, and high energy efficiency in massive MIMO or holographic MIMO in 6G.
\subsection{Multi-layer RIS}
While existing research primarily explores single-layer RIS, recent studies have highlighted the capacity of multi-layer RIS to perform equivalent communication functions with wireless signals. These functions encompass beamforming, physical-layer security, DOA estimation, and spectrum sensing. The transmission process between two RIS layers is shown in Fig.~\ref{background}(c), where the Rayleigh-Sommerfeld diffraction equation is adopted to quantify the propagation effect~\cite{Mengu2022MultilayerSR}. In other words, a signal transmitted through multi-layer RIS exhibits a linear or non-linear relationship with the input signal, and the relationship can be precisely defined through mathematical expression. Thus, the over-the-air computation capability of multi-layer RIS offers a scalable, light-computation-speed, energy-efficient alternative or complement to traditional communication functions.  
\subsection{Diffractional deep neuron networks (D$^2$NN) based on multi-layer RIS}
Beyond the potential applications previously discussed, prototypes of multi-layer RIS have been developed, demonstrating their capability to achieve deep neuron networks (DNNs)~\cite{Meta2022CuiNature}. Owing to the diffraction feature in enabling DNN, the architecture is aptly named D$^2$NN. The prototype depicted in Fig.~\ref{background}(d) demonstrates a wireless D$^2$NN for image transmission, with the first RIS layer modulating and transmitting the image as a RIS-MIMO transmitter. Then, three RIS layers form a D$^2$NN to decode the image, exemplifying the innovative application of multi-layer RIS in advanced signal processing and communication.

In summary, the relationship between the input signal and its transmission through multi-layer RIS, characterized by its mathematical expressibility, presents a scalable, high-speed, and energy-efficient alternative or complement to traditional communication technologies. Meanwhile, using RIS as a MIMO transmitter also presents a compelling solution for low-cost, low power consumption, and high energy efficiency massive MIMO or holographic MIMO in 6G.

\section{RIS-based diffractional channel coding (DCC)}
This paper presents RIS-based DCC, a signal-level coding scheme that utilizes the diffraction between two RIS layers. In this section, we elaborate on the system model, focusing on the diffractional encoder and decoder modules. We then introduce the basic principles of DCC, followed by examples of block codes and trellis codes. At last, the performance of the proposed scheme is demonstrated by comparing it to conventional channel coding schemes.

\subsection{System model}
\begin{figure}
\centering
\includegraphics[width=16.5cm]{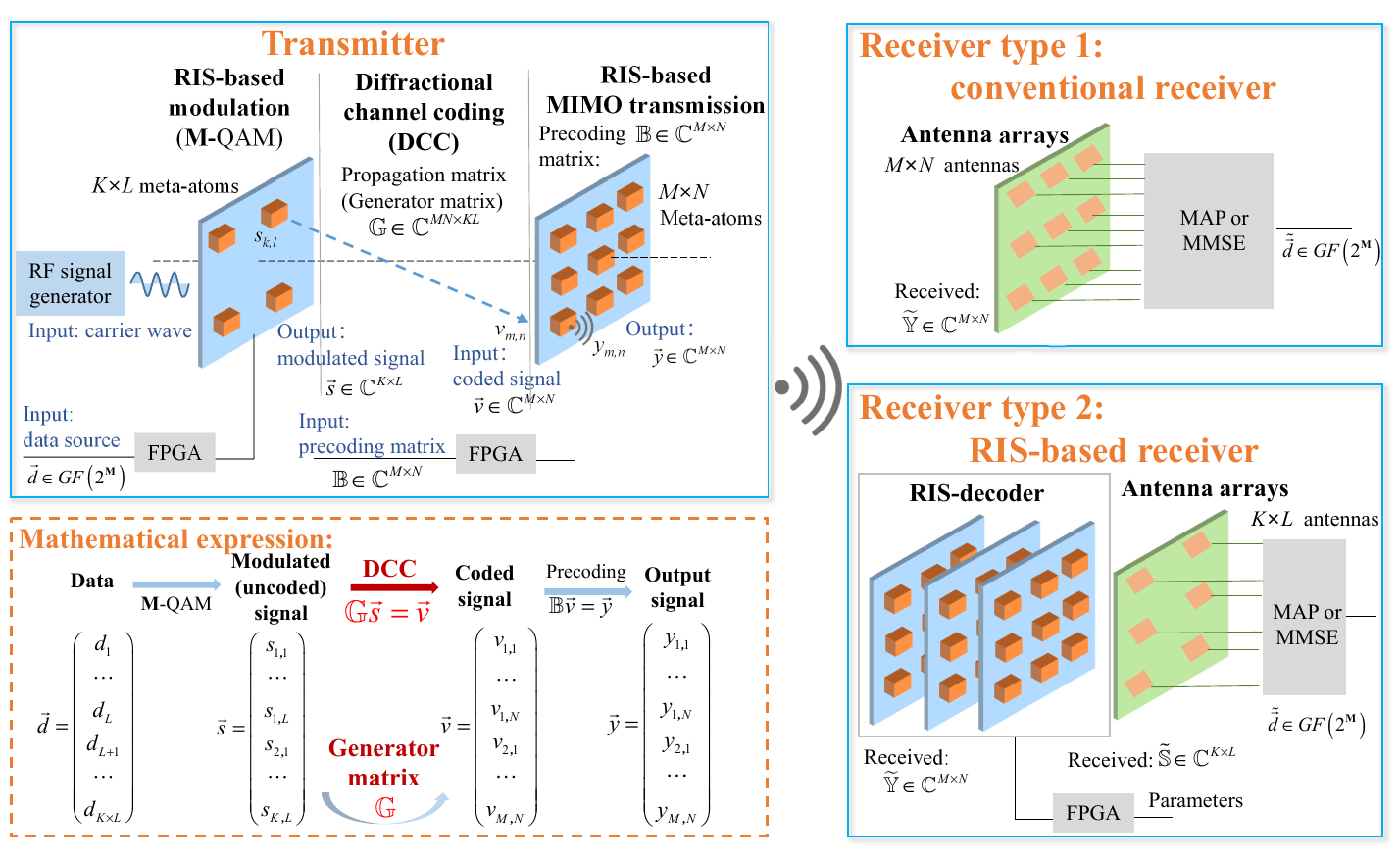}
\caption{Transceiver structure of RIS-based DCC, where two kinds of decoders, i.e., conventional and RIS-based decoder, are presented. Mathematical expressions within the transmitter are depicted. Similar to the conventional channel coding schemes, the dimension of the coded word in DCC is also larger than that of the uncoded word, and the coded word is the product of the uncoded word and the generator matrix. However, DCC operates directly on the signal, and modulation is required before DCC. }
\label{systemmodel}
\end{figure}

The RIS-based DCC system employs two layers of RIS for the integration of modulation, DCC, and MIMO transmission, as depicted in the transmitter in Fig.~\ref{systemmodel}. The system utilizes a first RIS layer with $K\times L$ meta-atoms and a second layer with $M\times N$ meta-atoms, where increased meta-atom count in the second layer enables signal dimension expansion after transmission between layers. Using the Rayleigh-Sommerfeld diffraction equation, we establish an equivalent generator matrix that captures the signal transformation from the first to the second layer. Meanwhile, the signal dimension expands from $K\times L$ to $M\times N$, and the additional transmitted signal is a combination of the input signal. Thus, the proposed scheme meets the basic requirements of channel coding and utilizes the property of diffraction between two layers, and we name it the DCC scheme.

Next, we will focus on the corresponding receiver. As shown in the first type of receiver in Fig.~\ref{systemmodel}, this setup is designed to ensure compatibility with the hardware of conventional MIMO systems. The core objective, i.e., to accurately retrieve information bits from the noise-added received signal, remains unchanged. However, in the proposed DCC method, the dimension of the information bit is smaller than that of the received signal. In other words, we need to recover fewer information bits from the received signal. Thus, the traditional signal detection techniques, including minimum-mean-squared-error (MMSE), orthogonal approximate message passing (OAMP), and DNN, could be effectively tailored to suit the DCC framework. Only modifications to the software are needed.

Meanwhile, inspired by the multi-layer RIS decoder in~\cite{Meta2022CuiNature}, a RIS-based MIMO receiver has been developed, i.e., Receiver type 2 in Fig.~\ref{systemmodel}. Multi-layer RIS forms a D$^2$NN so that the original information signal can be extracted. Then, utilizing conventional antenna arrays and an estimator, the system efficiently recovers the original information bits. In other words, the signal dimension is first reduced from $M\times N$ to $K\times L$ by D$^2$NN, aligning with the configuration of antenna arrays. Then, the estimator recovers the $K\times L$ bit streams from the corresponding signal streams. Thus, this type of receiver requires additional devices, i.e., multi-layer RIS, in front of the conventional MIMO receivers.

Beyond the structural differences, the two proposed decoders offer distinct benefits and limitations. The second receiver facilitates the offloading of complex signal detection operations from hardware to multi-layer RIS, which is particularly advantageous for future massive MIMO configurations. However, despite leveraging the inherent benefits of multi-layer RIS, the second structure exhibits less flexibility compared to the first one.

\subsection{Principles of DCC}
\begin{figure}
\centering
\includegraphics[width=10cm]{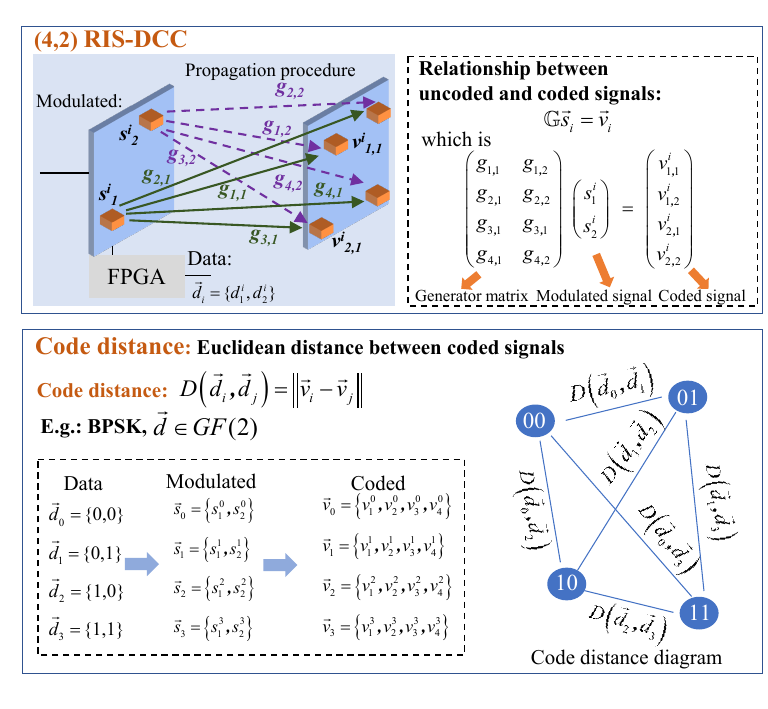}
\caption{Consider the example of (4,2) RIS-DCC to explain the generator matrix and code distance of the proposed scheme. There are 2 (or 4) meta-atoms in the first (or second) RIS layer. Data symbols are in the Galois field (2), and $\vec{d}_0$ to $\vec{d}_3$ are four datawords. Assuming BPSK, and $\vec{s}_0$ to $\vec{s}_3$ are four modulated and uncoded signals. $\vec{v}_0$ to $\vec{v}_3$ are four coded signals. The code distance is the Euclidean distance between coded signals in the complex domain.}
\label{principle}
\end{figure}

Although DCC employs a unique realization method compared to conventional channel coding schemes, it's crucial to analyze DCC within the established framework of traditional coding to highlight its compatibility and innovations. This subsection will detail the principles of DCC, such as the generator matrix, code rate, and code distance, illustrated through a straightforward example depicted in Fig.~\ref{principle}.

Consider an example where the first RIS layer comprises two meta-atoms and the second layer contains four, designed to illustrate DCC principles with block coding for an infinite data symbol stream\footnote{In practice, the number of meta-atoms significantly exceeds this example, which serves solely for explanatory purposes.}. Data streams are segmented into blocks of two data symbols (datawords), modulated by the first RIS layer. Assuming M-QAM modulation, symbols reside within the GF(2$^M$) field. After modulation, these two data symbols, which we refer to as uncoded signals, are encoded into a block of four coded signals (codewords) via DCC. These codewords are then analog-beamformed through the second RIS layer, forming a continuous stream of coded signals. This results in a code rate of 2/4, and thus we name this code (4,2) RIS-DCC. This rate reflects the ratio of meta-atoms between the second and first layers. As depicted in Fig.~\ref{principle}, the Rayleigh-Sommerfeld diffraction equation determines the signal propagation factor between two meta-atoms. Thus, the generator matrix, represented by the propagation matrix, multiplies the uncoded signal to produce the coded signal, aligning with traditional channel coding paradigms.

The code distance concept in DCC diverges from conventional channel coding, which relies on binary operations within the Galois field for distance calculation. Instead, DCC calculates code distance using Euclidean distance in the complex domain, given its direct operation on signal attributes like amplitude and phase. This distinction necessitates a unique approach to error detection and correction, as depicted in Fig.\ref{principle}. Unlike traditional methods, where errors are identified through bit flips, DCC addresses channel-induced amplitude and phase variations, emphasizing the identification of the most likely transmitted data symbols. The decoding spheres for DCC, illustrated in Fig.\ref{principle}, use complex number distances between codewords to measure code quality. Similarly, the minimum distance determines the code's capability, with larger minimum distances indicating better codes.

\subsection{Block and trellis DCC}
\begin{figure}
\centering
\includegraphics[width=16.5cm]{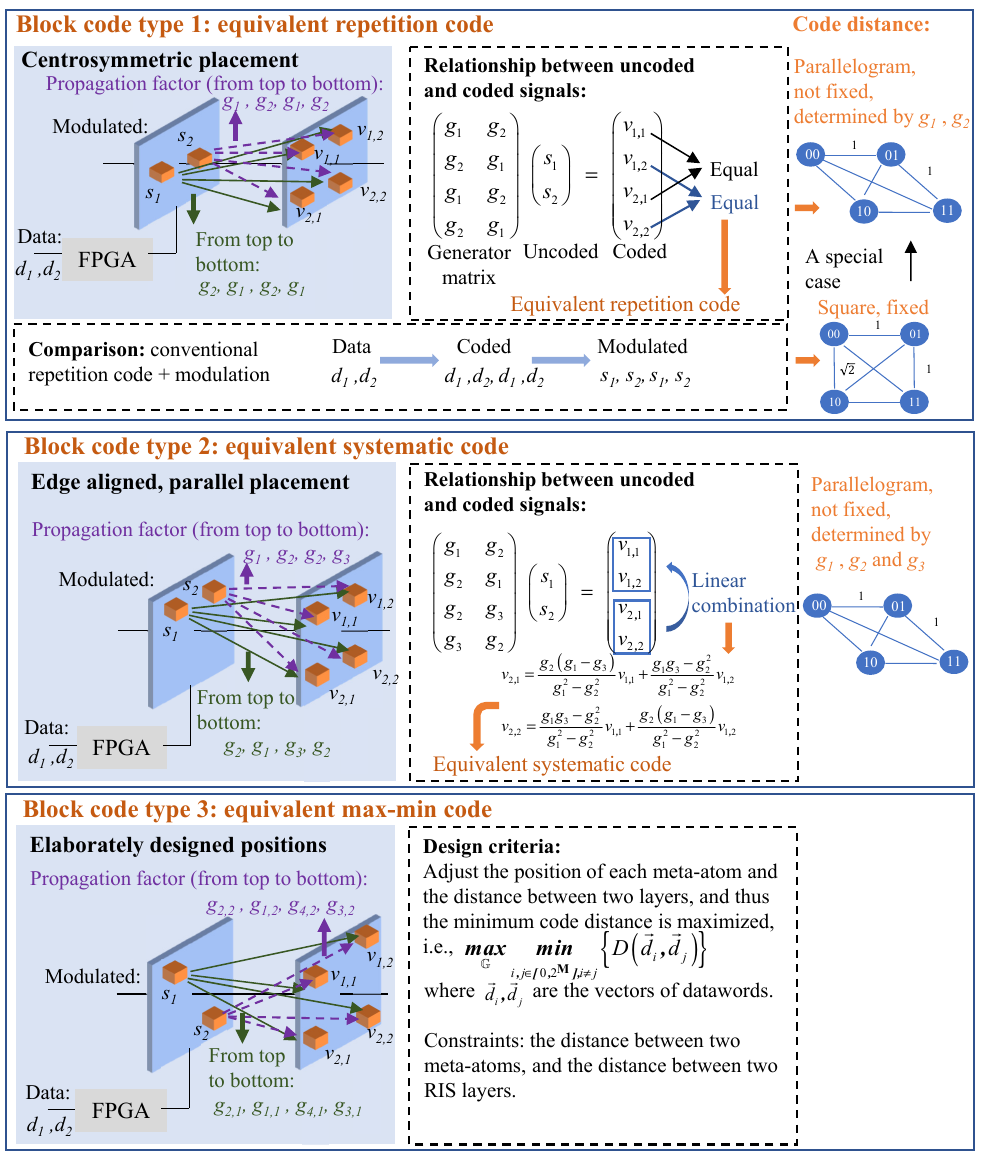}
\caption{There are three types of block codes, where the example of (4,2) RIS-DCC is adopted. $d_1$ and $d_2$ are the datawords. After modulation, $s_1$ and $s_2$ are the uncoded signals. $v_{1,1}$ to $v_{2,2}$ are the coded signals. }
\label{42RIS}
\end{figure}

\begin{figure}
\centering
\includegraphics[width=16cm]{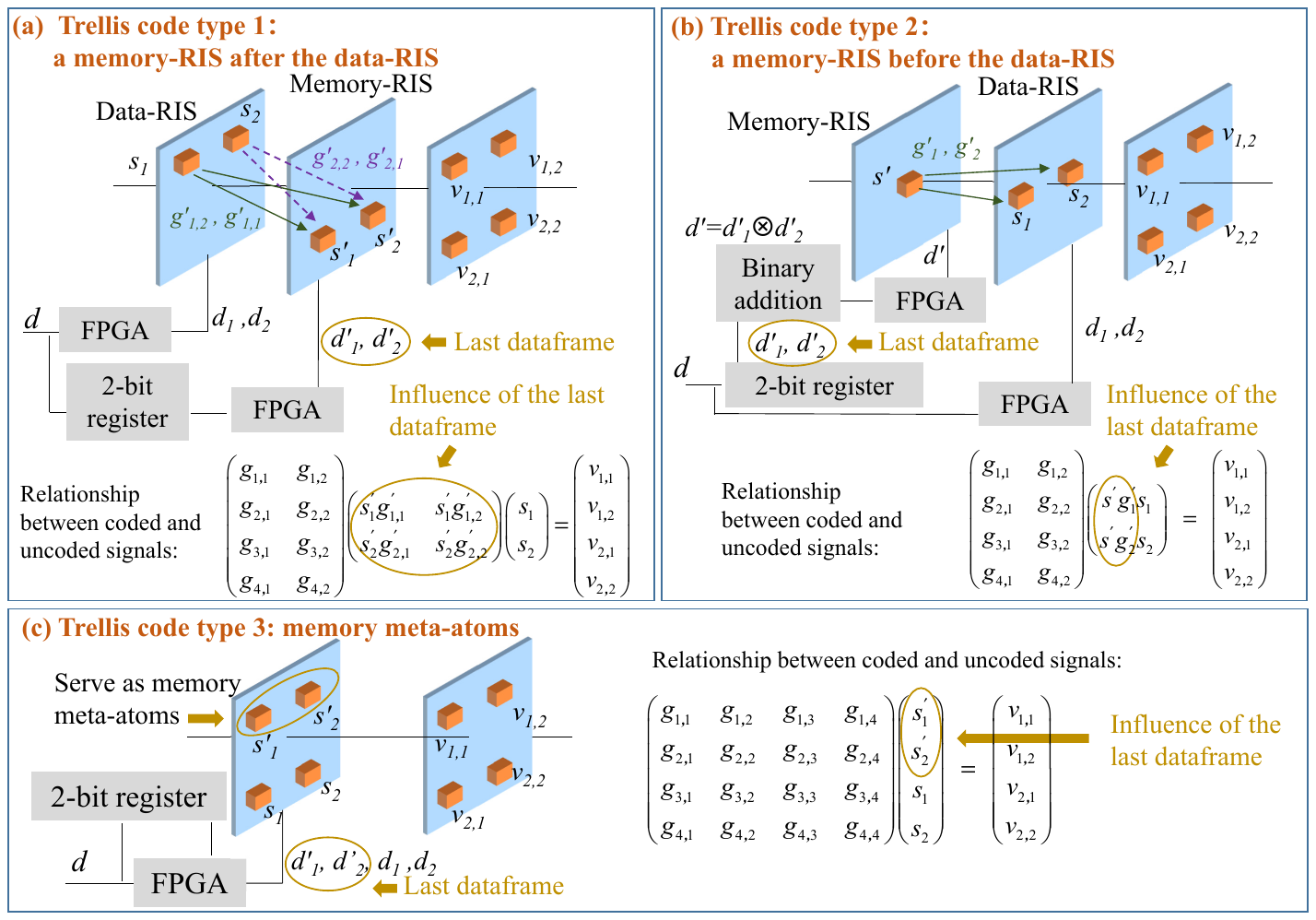}
\caption{There are three types of trellis codes. $d_1$ and $d_2$ are the current dataframes. $d_1'$ and $d_2'$ are the last dataframes. After modulation, $s_1$ and $s_2$ are the current uncoded signals, and $s_1'$ and $s_2'$ are the last uncoded signals. $v_{1,1}$ to $v_{2,2}$ are the coded signals. }
\label{memorycode}
\end{figure}

Traditional channel coding techniques are broadly categorized into block codes and trellis codes, differentiated by the encoding process's reliance on memory. This subsection explores the implementation of these established coding strategies within the DCC framework, as shown in Figs.~\ref{42RIS} and \ref{memorycode}, respectively.

Fig.~\ref{42RIS} illustrates the adaptation of three fundamental block codes to DCC, starting with the repetition code. By ensuring all meta-atoms are centrosymmetric and aligning two RIS layers parallel to each other, only two distinct propagation factors emerge, leading to the coded signal mirroring the first two signals. This alignment fulfills the criteria of repetition code, but with a different realization method. To be specific, DCC repeats the combination of modulated datawords, not the datawords themselves. This modification becomes evident when comparing code distances. Traditional repetition codes yield codewords that form a square with a constant distance between any two datawords. However, DCC forms a parallelogram, where the inner angles, i.e., code distances, adjust according to the meta-atom spacing and layer separation. Thus, DCC offers a flexible framework where the conventional repetition code can be seen as a special case of the proposed scheme.

Next, we move the meta-atoms on the first layer up to align with the meta-atoms in the second RIS layer, introducing what we term block code type 2, depicted in Fig.~\ref{42RIS}. There are three propagation factors, so the last two coded signals are a linear combination of the first two coded signals, which fulfills the requirements of systematic codes. Compared to conventional systematic codes, the proposed scheme uses a combination of coded signals instead of uncoded signals. The resulting code distance forms a distinct quadrilateral shape. By altering the configuration of the RIS layers, we can tailor the code distance to our requirements, showing the versatility of the proposed scheme for implementing systematic codes.

Inspired by the flexibility in adjusting code distances, we introduce a third block code variant, shown at the bottom of Fig.~\ref{42RIS}. Our strategy involves formulating a max-min optimization problem aimed at enlarging the minimum code distance. By adjusting the positions of the meta-atoms and the distance between the two RIS layers, we enhance system performance and achieve an optimal coding strategy.

Following the exploration of block codes within the DCC framework, we now turn our attention to how trellis codes can be implemented through multi-layer RIS, as shown in Fig.~\ref{memorycode}. For instance, two layers of RIS can be utilized to modulate the data, with one of the RIS layers acting as a memory RIS and the other acting as a data-RIS, as depicted in Fig.~\ref{memorycode}(a). The data-RIS operates under the current dataframe. The output signal of the data-RIS is the input signal of the memory-RIS, and the responses of the memory-RIS layer are controlled by the last dataframe. As a result, the coded signal not only contains the current dataframe but also encompasses the effects of the last dataframe. Alternatively, positioning the memory-RIS ahead of the data-RIS offers another solution, as depicted in Fig.~\ref{memorycode}(b). In this case, a binary addition is utilized to process the last dataframe, and the output of the binary addition controls the response of the memory-RIS. Thus, the influence of the last dataframe can also be observed in the coded signals. Additionally, Fig.~\ref{memorycode}(c) illustrates the integration of extra meta-atoms to directly achieve trellis coding. Both the last and current dataframes are utilized to control the response of the RIS. In this scenario, the relationship between the coded and uncoded signals is similar to that in traditional Trellis codes.


\subsection{Performance analysis}
\begin{figure}
\centering
\includegraphics[width=14cm]{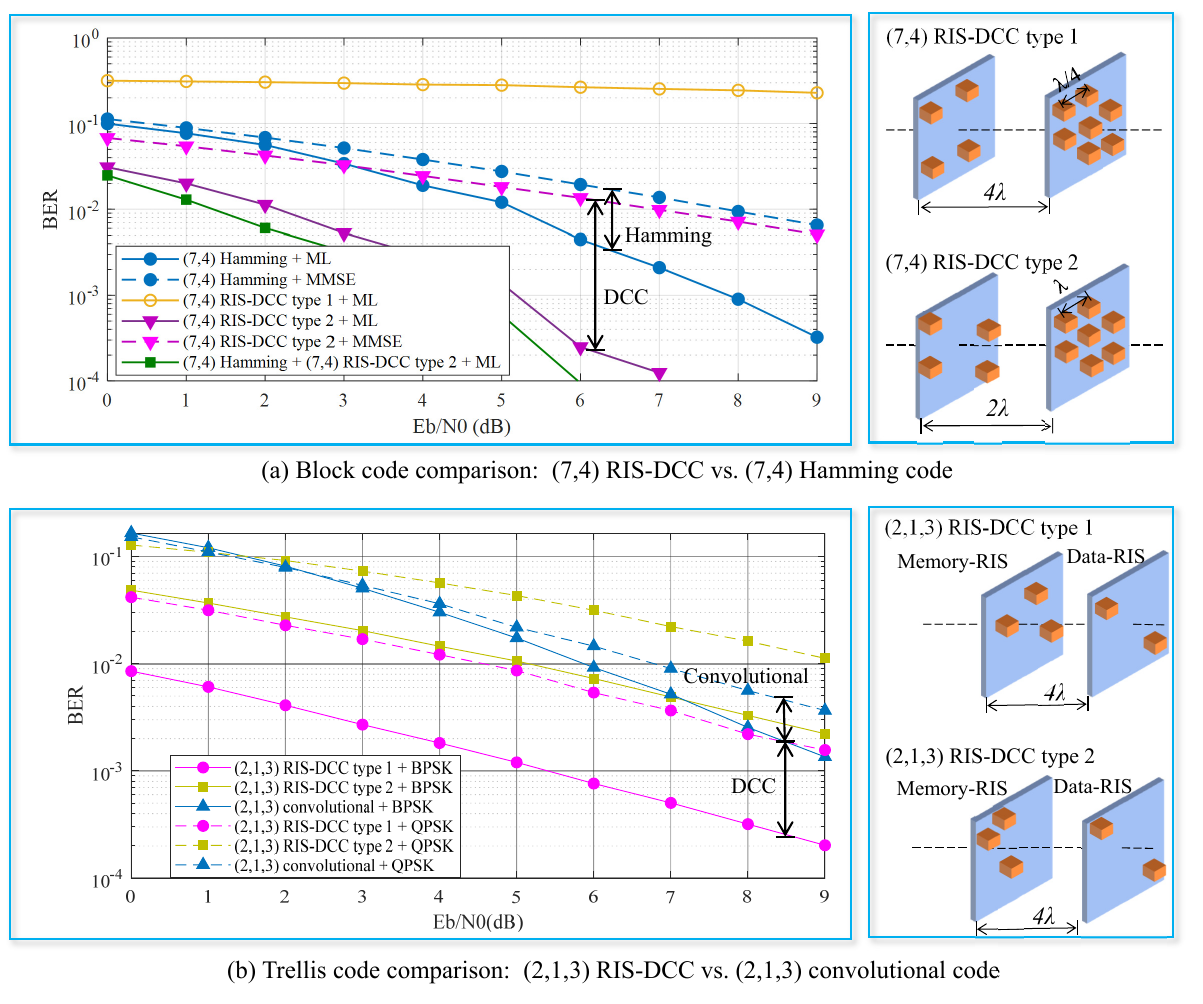}
\caption{Performance analysis of DCC, where two kinds of RIS structures are designed for (a) (7,4) RIS-DCC and (b) (2,1,3) RIS-DCC, respectively. In (a), MMSE and ML detectors are compared, and BPSK modulation is assumed. In (b), BPSK and QPSK modulations are compared, and the ML detector is assumed. }
\label{performance}
\end{figure}

The BER performance of the proposed DCC scheme is illustrated in Fig.~\ref{performance}, where BER is intrinsically linked to the physical configuration of RIS layers. For example, the BER of (7,4) RIS-DCC type 1 is higher than (7,4) Hamming code, while (7,4) RIS-DCC type 2 is lower than (7,4) Hamming code. Additionally, given its operation directly on the signal level, the DCC scheme can be seamlessly combined with existing channel coding strategies. As shown in Fig.~\ref{performance}(a), concatenating Hamming coding with RIS-DCC in sequence, represented by the combined scheme shown in green, exceeds that of each individual scheme. Furthermore, direct operation on the signal level makes DCC particularly sensitive to the selected detection and modulation strategies, shown in Figs.\ref{performance}(a) and (b), respectively. Thus, selecting or designing appropriate modulation and detection techniques, as well as the physical structure of RIS layers, is important to optimize the overall performance of the DCC scheme.

\section{Pros and cons of RIS-based DCC}
The RIS-based DCC approach brings several advantages to wireless communication systems, outlined as follows:
\begin{itemize}
\item\textbf{No power cost for encoding.} This scheme stands out for encoding signals without requiring additional power, utilizing natural signal propagation between RIS layers without extra hardware or energy.
\item \textbf{Computation at the speed of light}: The encoding process completes instantaneously as the signal propagates from one RIS layer to another, demonstrating unmatched computational speed. For instance, at millimeter-wave frequencies (e.g., 25 GHz), the time required for encoding can be as short as 0.4 ns.
\item \textbf{Compatibility with the current transceiver.} DCC can seamlessly integrate into existing transceiver systems. On the transmitting end, DCC can directly operate in RIS-based MIMO transmitters~[9] without requiring specialized hardware design. It can collaborate with other functions performed by RIS, such as modulation. On the receiving end, as demonstrated through two receiver models in Fig.~2, DCC aligns with existing MIMO systems, ensuring straightforward implementation.
\item\textbf{Versatility as a supplement or alternative.} By integrating extra signals for channel counteraction, DCC can either supplement or replace traditional coding schemes, potentially enhancing BER performance when used in combination with conventional schemes.
\item\textbf{Scalability with increasing signal dimension.}Traditional encoding methods face rapidly increasing computational complexity as the data dimension grows. In contrast, DCC leverages propagation, enabling processing at the speed of light regardless of the number of meta-atoms in each layer. This characteristic makes DCC suited for scenarios with large data volumes, massive MIMO, and holographic MIMO.
\item\textbf{Adjustable code distance.}The physical arrangement of RIS layers, i.e., meta-atom positions and layer distances, allows for flexible code distance adjustments to meet specific communication needs.
\item\textbf{Application potentialities in other wireless communication areas.} The function of one RIS layer can be regarded as a certain kind of signal transformation. It has been proven that different kinds of transformations, such as Fourier~\cite{Meta2020DingPhono}, Laplace transformation of spatial signal~\cite{Pan2021LaplacePhoto}, convolution of images~\cite{Fu2022ConvLight} and sequence signals, can be achieved by RIS. Thus, apart from signal-level coding, the proposed scheme has the potential to replace other signal processing modules in the future.
\end{itemize}

However, the DCC scheme also faces certain challenges.
\begin{itemize}
\item\textbf{Propagation expression accuracy.} Although the Rayleigh-Sommerfeld diffraction equation is widely used, it cannot accurately represent the propagation between two closely spaced RIS layers, such as when the interlayer distance is within 10$\lambda$. More precise expressions are crucial for accurate encoding and decoding design, such as full-wave methods or analytical electromagnetic models~\cite{mi2022}.
\item\textbf{Transmission power loss.} 
A notable consideration in DCC is the energy loss as signals pass
through RIS layers, and the amount of energy loss is influenced by various factors
like phase resolution, unit cell design, fabrication methods, frequency, and polarization~\cite{Wang2020PowerPRA}.  Therefore, the design of the DCC should carefully select the phase resolution, design the transmission structure, and consider system performance trade-offs. Fortunately, DCC can flexibly use 1-bit transmissive RIS, which has a transmission energy loss as low as 0.2 dB.
\item\textbf{Physical structure limitation.} Although we can alter the physical structure of RIS to design the codes, there are limitations induced by the structure itself, such as the distance between two meta-atoms and the distance between two RIS layers. For example, the distance between two meta-atoms cannot be too large or too small. In~\cite{Tang2020RISJSAC}, the authors limited the distance between two meta-atoms to the range of $[\frac{\lambda}{10},\frac{\lambda}{2}]$. Therefore, the design of DCC should meet the physical requirements of RIS.
\item \textbf{Operational challenge}. In practical deployment, DCC may face operational challenges, including channel estimation, impedance matching, and mutual coupling elimination. Although these challenges are not the focus of this paper, they are crucial as they affect the overall efficiency and performance of the system.
\end{itemize}

\vspace{0.25in}
\section{CONCLUSION}
In this paper, we introduce a pioneering signal-level channel coding scheme for wireless communication systems, termed RIS-based DCC. By harnessing the diffraction of multi-layer RIS, our scheme offers a novel approach to encoding signals. We detailed the system model and principles of DCC, including the generator matrix, code rate, and code distance, and illustrated the implementation of block and trellis codes within the RIS-DCC framework.

The scheme presents multiple advantages. For example, it requires no additional energy for encoding, achieves computation at light speed, and offers flexibility in code distance adjustment. This paper also highlights the potential of DCC to either replace or complement existing channel coding methods. However, as an emerging concept, RIS-based DCC unveils numerous research opportunities, demanding further investigation into its theoretical foundations and practical applications. Key areas identified for further investigation include:
\begin{itemize}
\item\textbf{DCC theory.} Establishing a robust theoretical framework for DCC, especially considering its operations in the complex domain, is crucial. This involves redefining traditional coding metrics and principles to suit the complex nature of signal propagation between and through RIS.

\item\textbf{Other equivalent coding schemes.} We also need to broaden the scope of DCC to either match or surpass existing coding protocols, including LDPC and Turbo codes. It requires not only mapping these established codes into the DCC domain but also exploring the potential for novel coding schemes that DCC's unique properties might enable.

\item\textbf{Decoding methods.} The development of practical and low-complexity decoding algorithms remains a challenge within traditional channel coding schemes. Given its operation at the signal level, DCC introduces additional complexities in designing decoding methods. This task requires a sophisticated blend of signal detection and conventional channel decoding techniques, presenting unique challenges that necessitate innovative solutions.

\item \textbf{Explainable theoretical model.} Considering that DCC relies on the physical structure of RIS and performs encoding during signal propagation, we need to establish an explainable theoretical model for DCC. This task requires integrating electromagnetic information theory and antenna-circuit theory to accurately characterize it.

\item \textbf{Integrate with other field.} 
Incorporating DCC into designs in other fields would be highly beneficial. The inherent MIMO transmission structure of RIS promotes the integration of DCC with massive MIMO. This synergy makes the effective coordination of DCC with other functionalities of MIMO communication systems a crucial focus for future developments. Furthermore, the miniaturized architecture of multi-layer RIS allows DCC to assist IoT devices in overcoming channel effects in space-constrained scenarios, ensuring highly stable communications.
\end{itemize}
\vspace{0.25in}

%

\vspace{0.25in}
\balance

\vfill

\end{document}